\documentclass[a4paper,showkeys,floatfix,aps,pre,reprint,groupedaddress]{revtex4-1}
\usepackage{amsmath,amssymb}
\usepackage{graphics,graphicx}
\usepackage{dcolumn,bm}
\usepackage{psfrag}
\usepackage{color}
\newcommand{\srmphysics}
{\affiliation{Department of Physics, SRM University - AP,
Andhra Pradesh - 522502, India}}
\newcommand{\srmcse}
{\affiliation{Department of Computer Science and Engineering, SRM University - AP, Andhra Pradesh - 522502, India}}

\begin{document}
\title{Machine learning predictions of COVID-19 second wave end-times in Indian states}

\author{Anvesh Reddy, Hanesh Koganti, Sai Krishna, Suhas Reddy}
\srmcse
\author{Soumyajyoti Biswas}
\email{soumyajyoti.b@srmap.edu.in}
\srmphysics
\begin{abstract}
The estimate of the remaining time of an ongoing wave of epidemic spreading is a critical issue. Due to the variations of a wide range of 
parameters in an epidemic, for simple models such as Susceptible-Infected-Removed (SIR) model, it is difficult to estimate such a 
time scale. On the other hand, multidimensional data with a large set attributes are precisely what one can use in statistical learning 
algorithms to make predictions. Here we show, how the predictability of the SIR model changes with various parameters using a supervised learning
algorithm. We then estimate the condition in which the model gives the least error in predicting the duration of the first wave of the COVID-19
pandemic in different states in India. Finally, we use the SIR model with the above mentioned optimal conditions to generate a training data set
and use it in the supervised learning algorithm to estimate the end-time of the ongoing second wave of the pandemic in different states in India. 
\end{abstract}


\maketitle
\section{Introduction}
Since the outbreak of the COVID-19 pandemic \cite{who}, the estimate of
a time-scale for the end of a wave of pandemic outbreak has undoubtedly become an outstanding challenge. Nevertheless, due
to the variations of a wide range of parameters, such as the rate of spreading, the contact network of the individuals, various mitigation measures etc., 
it is very difficult to make such an estimate \cite{chin,covid5,covid6,covid7,covid8,covid9,ps2,covid3}. However, 
a multidimensional set of data is often used in statistical learning approaches
for making predictions \cite{leo,ml_book}. Indeed, such predictions have been attempted in a wide range of cases, such as financial time series, weather data, medical applications
and many other physical systems \cite{labquake,salm18,geophys,sb}. There have been multiple earlier attempts in using machine learning approaches for predicting epidemic spreading in the context of COVID-19 (see e.g., \cite{ai_covid1,ai_covid2}) as well. However, to have a proper estimate, a large set of training data is
 needed to be fed to the supervised learning algorithm. 
This is often a major hurdle to overcome for a pandemic such as the present one, the like of which is not seen in a century.

To address this issue, we first consider a simplified model of epidemic spreading, called the Susceptible-Infected-Removed (SIR) model \cite{sir_o,ep1,ep2} and 
estimate its predictability using a supervised learning algorithm, by varying various parameters of the SIR model. We then find the condition under which 
the model is best suited to make `predictions' about the first wave of the COVID-19 pandemic in different states in India. Since in most of these cases,
the first and the second waves are separated by a period of low infection rates, the end-time of the first wave can be well defined. Therefore, it is 
possible to make an error estimate for the `prediction' of the first wave end-time. The optimal condition of the model that `predicts' the end of the
first wave can then be used to generate a `synthetic' training data set of substantial size using the simulation data of the SIR model. This `synthetic' training data 
set can then be used 
to make predictions for the ongoing second wave. The use of synthetic data for enhancing
prediction capability of ML algorithms is a well known technique (see e.g., \cite{syn}). By increasing the size of the training data set substantially, this technique enables 
the ML algorithm to make stable predictions. 

Certainly, there are multiple issues in using the first-wave data for the optimization of the training set.
Particularly,  the two wave are, of course, different in several aspects: effects of vaccinations, changes in the norms of travel restrictions, presence of mutant variants of the virus, etc. One outcome of these variations would be the changes in the maximum values of the daily infection rates between the two waves. As is evident from 
the data in India \cite{data}, the peak height of the second wave was about four times larger than the peak height of the first wave.
Therefore, we normalize the data by the peak height, which
necessarily assumes that the peak for the second wave has passed, in each of the cases where we make the end-time predictions.   

The rest of the paper is organized as follows: First we describe the SIR model and the machine learning methods used for making the predictions in the model (sec. II). 
In sec. III we present the simulation results, describing the variations in predictability of the SIR model under different conditions (testing rate, site dilution). Then
in sec. IV we use the ML algorithm to make predictions of the end-time of the second wave of the pandemic for some states in India. Finally we discuss and conclude in sec. V.

\section{Model and methods}
The SIR model is a well studied model for epidemic spreading \cite{sir_o}. Although a simple model, this and its variants have been widely popular for epidemic spreading 
studies \cite{ep1,ep2}.
The model assumes the total population to be divided into three groups - Susceptible: denoting the individuals who can get infected by the virus but are not yet
infected, Infected: denoting the individuals who are currently infected and can infect the susceptible population and Removed: denoting the population 
who were already infected by the virus and do not affect the evolution of the spreading dynamics any more. 

We consider the model on a two dimensional square lattice, where each site represents the location of an individual who can be in one of the three states mentioned above. 
An infected individual can, with certain probability, infect one of their eight neighbors (four nearest neighbors and four diagonal neighbors) if that neighbor
is in the susceptible state. An infected individual remains in that state for a given duration of time, during which they can infect other susceptible individuals. 
Following that duration, the infected individual enters the Removed state, where they no longer participate in the spreading dynamics.

In one time step of the simulation, every individual is selected once and their states are attempted for a possible update. 
The updates are done in a parallel updating scheme, such that
a given update comes into effect in the following time step. If an individual is in the susceptible state and one of the eight neighbors is infected, then the
susceptible individual can be infected with probability $p$. If an infected individual has remained in that state for $\tau$ time steps, then that individual is put
in the removed state. Here we keep $\tau=14$ and the value of $p$ is fixed at a randomly chosen value between 0.3 to 0.8 from a uniform distribution for each realization of the model.

The number of infected individuals at a time $t$, denoted by $I(t)$, represents what is usually refereed to as the `active cases'. The time derivative of 
the susceptible ($S(t)$) individuals, $dS(t)/dt$, is essentially the number of new infections in a day (at $t$). Both of these quantities start from 
a low value, representing the initial infected individuals, which is chosen randomly and uniformly between 10 and 20 for each simulation. Both of these
quantities then increase with time, reach a peak and then eventually decrease to zero. The model does not show multiple waves of infection rates, nor does
it account for the effects of vaccination or restriction imposed in interactions. 

\begin{figure}
\includegraphics[width=8cm]{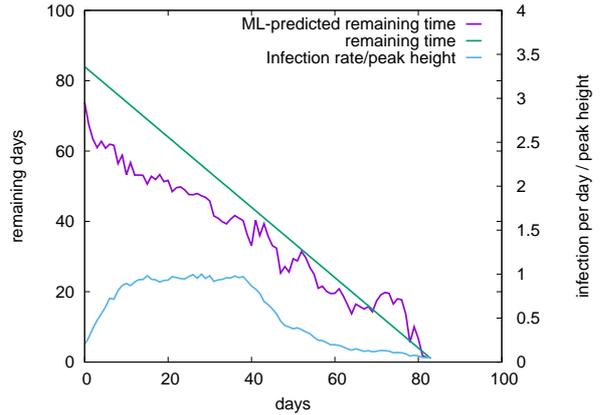}
\caption{The time variation of the daily infection rate (normalized) and ML predictions for the time remaining before the
end of infections are shown for a single run of the SIR model. The actual remaining time (straight line) is also shown, and
the root-mean-square deviation between the predicted and the actual time gives the error in the prediction.}
\label{single_prediction}
\end{figure}

While it is straightforward to get an exact solution for the mean field version of the model and also to numerically estimate the above mentioned quantities in
other topologies (including the square lattice considered here), the actual situation is far more complex and the available data sets are limited. Particularly, just the 
absence of tests for the individuals without symptoms and/or access to such medical facilities, would distort the data for the number of infections and other related
quantities. Also, the topology of a square lattice is a simplified one and would at the very least include `disorder' in terms of unoccupied sites. We first
investigate the effects of these factors in predictability of the SIR model. For the prediction, we use a supervised machine learning algorithm, particularly
the Random Forest algorithm. This is an ensemble of decision tress and the predictions in the model are made using the majority of the predictions of the decision trees.
The various attributes that we use for the training of the algorithm are: daily infection, daily recovery and the number of active cases at a particular time. The target variable
for the prediction is the remaining time before the daily infection number goes below 5\% of the peak value. In the Random-forest, we used 1000 estimators and kept the maximum depth at 15. The results are stable with small variations around these parameters. Following the training of the algorithm with a training set of 
200 ensembles (each ensemble represents the full time series of the different quantities mentioned from the start to the end of the spreading dynamics), each having
a randomly chosen infection rate and initial infection number, in the way outlined above. Then the trained algorithm is used to make predictions for a different set of 
100 ensembles. 
\begin{figure*}
\includegraphics[width=17cm]{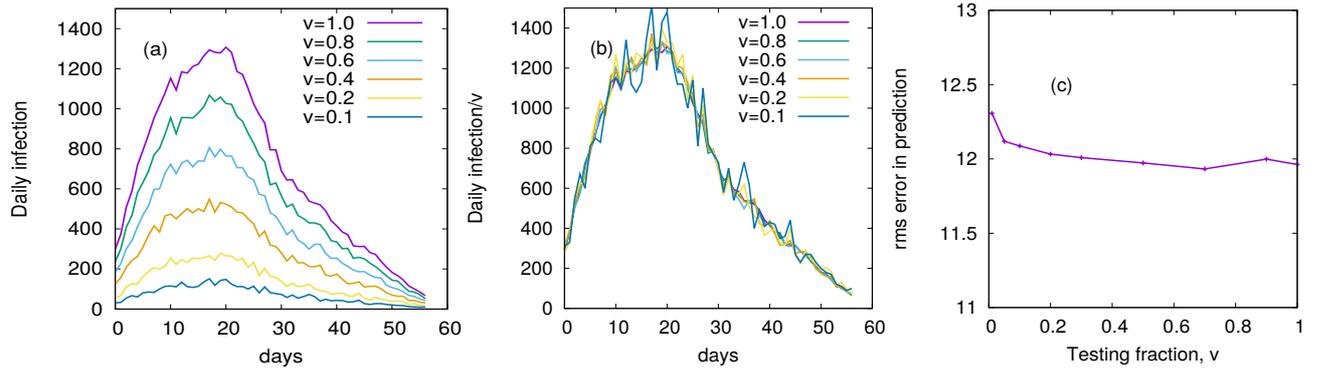}
\caption{The simulation results for the SIR model in two dimensional square lattice ($200 \times 200$) for a fixed
fraction $v$ of the population tested. (a) The time variation of the daily infection rate for a particular realization
of the model is shown for different values of $v$. (b) When scaled by $v$, the daily infection curved fall on top
of each other, with a variation in fluctuations. (c) The variation in the root-mean-squared error in prediction made by the ML
algorithm (for 100 sets, after training by 200 sets) with $v$ is shown. No significant variation is noted.}
\label{variabletest_combined}
\end{figure*}
In Fig. \ref{single_prediction}, a typical time series of the infection rate (normalized by the peak height) is shown. The actual
remaining time and the machine learning (ML) predicted remaining times, at every instance, are also shown. The root-mean-squared
fluctuations between the actual remaining time and the predicted remaining time at every instance, gives an estimate for the error
in prediction. In the following, we first estimate the error in predictions i.e., the efficiency of the ML predictions under
different conditions of variable testing rate and disorder in the SIR model. Then we use the model as training set to make 
predictions for the end of the second wave of the COVID-19 pandemic in different states in India.

\section{Simulation results}
Here we describe the simulation results of the SIR model of epidemic spreading and estimate the variations of its predictability with different parameters of the model, using supervised machine learning algorithm. 
\subsection{ML predictability of SIR model with variable testing}
As indicated before, a major source of distortion in the data for the pandemic is the limited testing resources
available. This was especially apparent during the first wave of the pandemic (see e.g., \cite{testing}). Therefore, it is useful to
understand, even for this simplified model, how does incomplete testing and/or variable testing rate affect the 
measurements in the model so as to affect, in turn, the predictability of the model. 

We check this effect in two different ways. First we assume that only a fraction ($v$) of the total population can get
tested. While the underlying SIR dynamics runs with the three possible states ($S$, $I$ and $R$) for each individual, 
the measurements are made, at each time, only on the $v$ fraction of the total population, chosen randomly and kept fixed in time for that realization.

\begin{figure*}
\includegraphics[width=17cm]{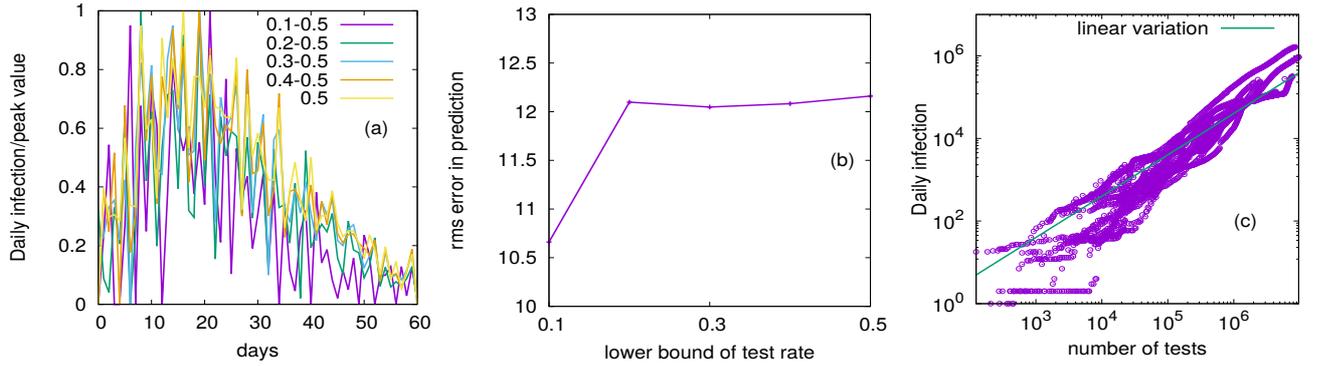}
\caption{The simulation results for the SIR model in two dimensional square lattice ($200 \times 200$) for a
time dependent testing rate are shown. (a) The time variation of the daily infections (scaled by the peak value)
are shown for various ranges of allowed variations in $v$. The curves are of different fluctuations, but of
similar nature.  (b) The variations in the error in ML predictions with the lower bound in testing fraction ($v_{min}$),
showing no systematic variation in predictability. (c) The actual dependence of the testing and daily positive cases are shown for different states
in India that shows an approximate linear variation (indicated by the straight line for a guide to the eye). This 
is for the justification of the choice of the linear dependence of $v$ with daily (apparent) infection rate 
in the simulations.}
\label{testrate_variation_combined}
\end{figure*}
Fig. \ref{variabletest_combined} depicts the effect of the various values of $v$, between 10\% ($v=0.1$) to 100\% ($v=1$)
testing. While the (apparent) number of daily infection is very sensitive to the value of $v$, when scaled by this
factor (Fig. \ref{variabletest_combined}(b)), the curves fall on top of each other, with a varying degree of fluctuations. 
The consequent error in the predictions using the ML algorithm, however, only varies weakly (Fig. \ref{variabletest_combined}(c)) 
with $v$. This gives an important conclusion that while drastically sub-sampling, the predictability remains almost the same in 
the model, as long as a macroscopic fraction of the randomly chosen population is tested.

Secondly, it is also known that the rate of testing is not a fixed quantity over the duration of the pandemic. Particularly, 
it is often dependent on the rate of positive results obtained in daily testing. Therefore, we also look at the variation due
to a time dependent value of $v$. We vary $v$ between a lower bound $v_{min}$ and an upper bound $v_{max}=0.5$, linearly 
dependent on the daily infection rate. Other than the linear dependence of $v$ within this range, it is not allowed to fall
below or increase above, the fixed threshold values. The individuals are again randomly selected for testing at each step, but
now with a time dependent value of $v$. Fig. \ref{testrate_variation_combined} shows the results for this case. In 
Fig. \ref{testrate_variation_combined}(a), the time variation of the daily infections are shown for various values of 
$v_{min}$ and $v_{max}$. In Fig. \ref{testrate_variation_combined}(b) the actual data for number of testing and the corresponding daily
infections are shown for various states in India, justifying the choice of the linear variation (indicated by the straight line). 
Nevertheless, there is hardly any systematic variation in the error (hence also the predictability) with $v_{min}$. 
This is also an interesting observation that while the time dependent testing rate can introduce fluctuations in the data, 
ultimately it does not translate to a lower predictability, as long as it is made sure that a macroscopic fraction of the individuals are
always tested. 
\begin{figure*}[tbh]
\includegraphics[width=14cm]{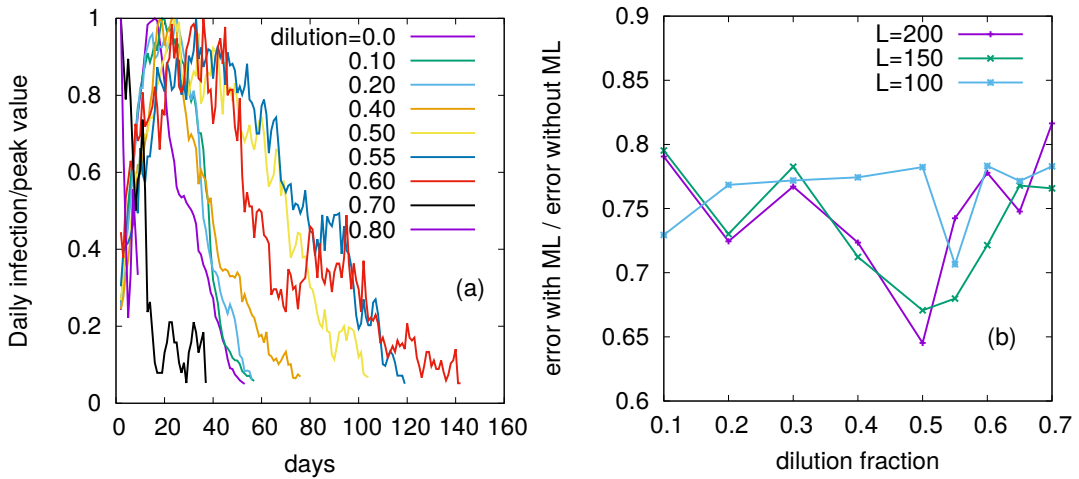}
\caption{The simulation results for the SIR model in two dimensional square lattice ($200 \times 200$) for different
values of site dilution fraction. (a) The duration for which the daily infection rate is finite, varies non-monotonically
with site dilution. Therefore, the curves, even when scaled, do not overlap with each other. (b) The relative errors for various
system sizes shows a minimum that tends towards the percolation threshold with increasing system size.}
\label{dilute_combined}
\end{figure*}

\subsection{ML predictability of SIR model with site dilution}
As mentioned before, topology of the contact network of the individuals can play a crucial role in the spreading dynamics. 
So far we have kept that to be very simple fully occupied two dimensional square lattice. But such an orderly arrangement 
is not realistic. As a simple way to introduce disorder, we remove a fraction $q$ of the sites i.e., there are no individuals
occupying that fraction of the sites
This modification, of course, introduces a fluctuation that diverges near the critical point $q_c\approx 0.4$ of site percolation \cite{stau} (see also
\cite{galam} for percolation threshold with longer than nearest neighbor connections). 
It is generally known that a system with higher disorder is relatively more predictable through machine learning, compared to the
systems having less disorder \cite{sb}. It is also known that the distribution of population in a city follow a fractal character \cite{city},
which will happen here near the percolation threshold. It is, therefore, interesting to study the variation of the predictability when
the SIR model is simulated in a site diluted lattice.

Fig. \ref{dilute_combined} shows the simulation results for the site diluted lattice. It is interesting to note that even when the infection curves are 
scaled by the corresponding maximum values, they do not overlap. Indeed, there is a non-monotonic variation in the duration upto
which the dynamics run, with the dilution fraction. The corresponding errors, scaled by the error obtained without ML i.e., just considering the
average duration of the training set as the predicted duration for each testing set, show a non-monotonic variation with
the dilution fraction. A system size dependence shows that the minimum point of the error tends towards the critical percolation
threshold, as the system size is increased. Therefore, we conjecture that the highest disorder in the model i.e., the percolation critical
point, is the maximally predictable point as well. This is an interesting observation, since as mentioned bore, at the
percolation point, the occupied site form a fractal structure. As mentioned bore, this mimics the fractal nature of
the population distributions in cities \cite{city}, although with a different fractal dimension.

\section{Application: Predictions of end-time of second wave in some Indian states}
So far we have discussed the predictability of the SIR model using supervised machine learning. We have also seen that
the predictability depends on the site dilution fraction in the model when simulated on a square lattice. Here we attempt 
in using the SIR model as a training set and then make predictions for the end-time of the second waves in eight Indian states
where the total infection has crossed one million. These states are: Andhra Pradesh (AP), Delhi (DL), Karnataka (KA), Kerala (KL),
Maharashtra (MH), Tamil Nadu (TN), Uttar Pradesh (UP) and West Bengal (WB). 
\begin{figure}[tbh]
\includegraphics[width=8cm]{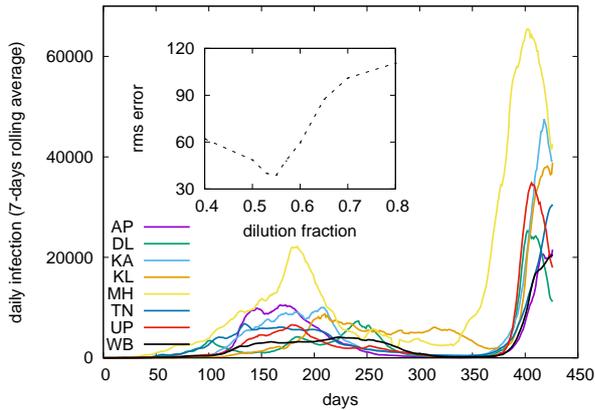}
\caption{The seven day rolling average of the daily infection from March 14, 2020 to May 20, 2021 in eight Indian states. The inset shows the
errors in `predicting' the first wave using training data from SIR model with various levels of site dilution. The minimum error is noted for the dilution
fraction 0.55, which is then used as the training set for the predictions of the second wave.}
\label{eight_states_roll}
\end{figure}

In Fig. \ref{eight_states_roll} we see that the first and second waves are more or less distinct in these states - separated by a low
daily infection rate. First we use the SIR model with various site dilution fractions and make `predictions' about the end-time
of the first wave. Knowing the actual end-time of the first waves in these states, it is possible to estimate the errors 
in those predictions (Fig. \ref{eight_states_roll}(inset)). It is seen that the error is minimum for the dilution fraction 0.55. 
We therefore use the SIR model with dilution fraction 0.55 to generate a training set (500 sets) and then feed the data for the
second wave to make a prediction about the end-time. In doing so, one obvious issue is with the peak height, which are very much different
between the first and the second waves and also among the different states. We, therefore, normalize the training as well as the testing data
by the corresponding peak heights. One obvious assumption, therefore, is that the peak for the second wave has passed, which is 
obvious in many of the states and are also indicative in the rest of the states. 

\begin{figure*}[tbh]
\includegraphics[width=14cm]{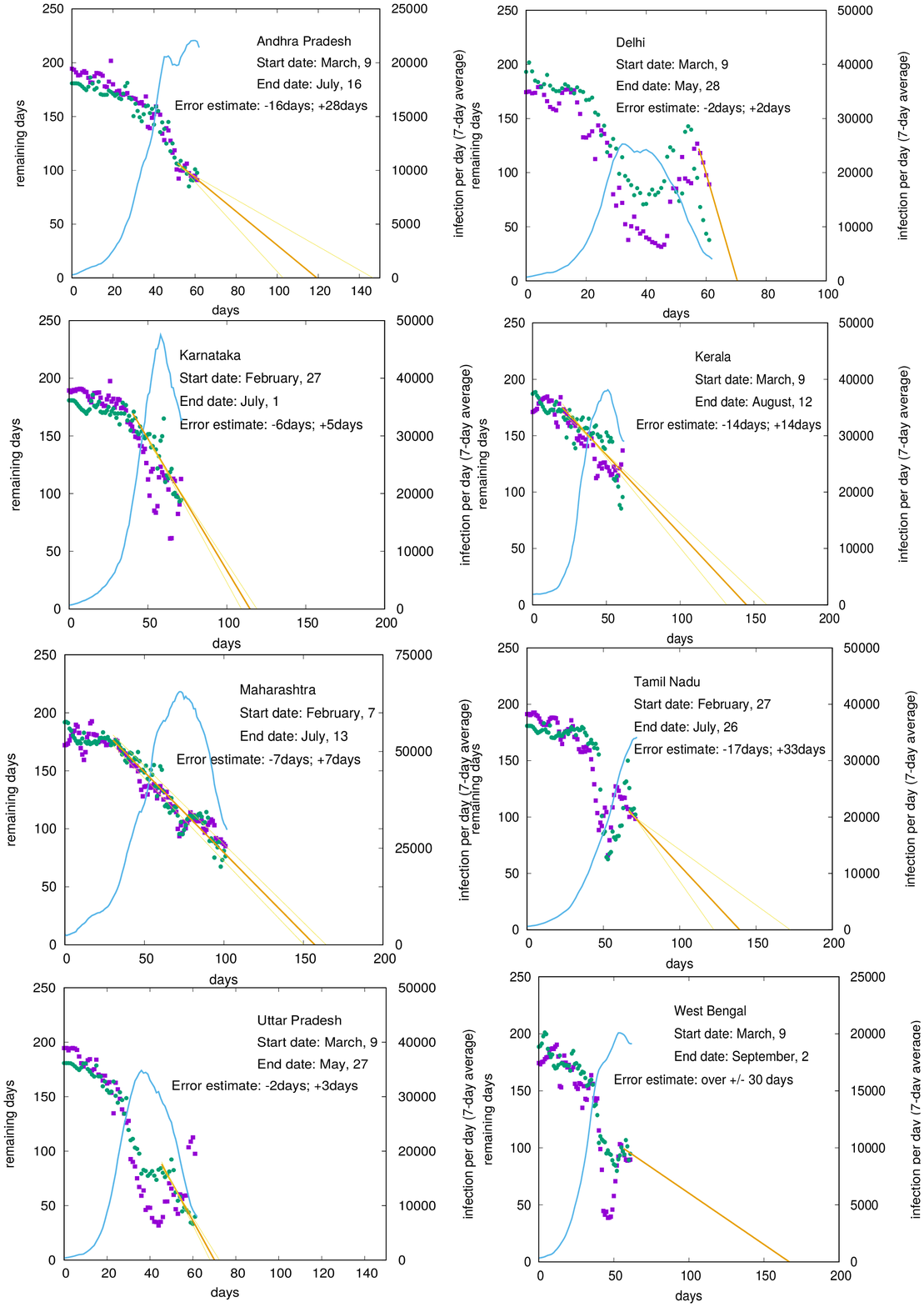}
\caption{The predicted end-time for the second wave in the eight Indian states, where the total infection has crossed one million. The errors are
also indicated, which are estimated from the least squared fit of the straight lines of the final points of prediction. In cases where the infection
rates are very close to the peak, the predictions are less accurate. The green dots are the predictions using SIR model as training set and the purple
dots are the predictions using the first wave data as the training set.}
\label{all_states_predictions}
\end{figure*}

We make another set of predictions by using the first waves as the training data. It is remarkable that the two sets of predictions are 
very close to each other. However, in making the final prediction (Fig. \ref{all_states_predictions}, Table 1), we use the training set of the SIR model.

\begin{table*}[t]
\centering
\begin{tabular} { | p {1.7 cm} | p {1.7 cm} | p {1.7 cm} | p {1.7 cm} | p {1.7 cm} | p {1.8 cm} | p {1.7 cm} | p {1.7 cm} | p {1.7 cm} |}
\hline
States & Andhra Pradesh & Delhi & Karnataka & Kerala & Maharashtra & Tamil Nadu & Uttar Pradesh & West Bengal \\
\hline
End-time & July 16 & May 28 & July 1 & August 12 & July 13 & July 26 & May 27 & September 2 \\
\hline
Errors & -16 days,+28 days & - 2 days, +2 days & -6 days, +5 days & -14 days, +14 days & -7 days, +7 days & -17 days , + 33 days &  -2days, + 3 days & -30 days, + 30days\\
\hline
\end{tabular}
\label{table_1}
\caption{The predicted end-dates for the second wave in different states and the corresponding errors. The errors are higher for the states where the infection rates are close to the peak (data as of May 20, 2021 \cite{data}).}
\end{table*}
\section{Discussions and Conclusions}
We have reported the variations in the predictability of the SIR model of epidemic spreading with different parameters of the model, using supervised machine learning algorithm. It is interesting to note that the predictions for the end-time in the model is remarkably stable, even when only a small fraction (10\%) of the individuals are tested for the infection. The predictions are also stable when the testing rate vary with time - linearly with the positivity rate of the testing within a given range. However, the predictability changes substantially when a disorder is introduced in terms of site dilution in the model i.e., some positions are not occupied by any individual. Particularly, the relative predictability of the model is the highest (error in prediction is the lowest) when the site dilution fraction approaches the percolation critical point (see Fig. \ref{dilute_combined}). In that case, the underlying lattice structure approaches a fractal and the fluctuations in the cluster size diverges with system size \cite{stau}. It is seen before that the predictability using ML approaches increases with the increase in the disorder in the system (see e.g., \cite{sb}). Indeed, the fluctuations in the time series of the various attributes used for the ML algorithm have richer characteristics and consequently the training of the algorithm is better. Also, it is interesting to note that the 
spatial distribution of population in cities are fractal in nature \cite{city}, although not necessarily of the same fractal dimension as that of the site percolation. Nevertheless, the fluctuations introduced in the daily infection rate and other related quantities due to the delayed spreading of the infections in marginally connection regions, would introduce qualitatively similar effects in any fractal geometry. 

We then use the model and the ML approach to make predictions of the end-time of the ongoing second wave of the COVID-19 pandemic in eight Indian states, where the total number of infections are over one million (see Table 1). In doing so, we first need to overcome the lack of training data for the ML algorithm. We first note that the first and second waves of the pandemic in India are somewhat separated by a relatively low daily infection rates. Therefore, we take the data for the first wave and make `predictions' about its end-time using the SIR model as the training set. As the predictability of the SIR model is already shown to be sensitive to the lattice dilution, we estimate the dilution fraction for which the error in the `predictions' for the first wave is minimum. We then use the SIR model with that dilution fraction to generate the training data set for making predictions of the end-time of the ongoing second wave. This approach assumes that the statistical nature of the fluctuations in the first and second waves would be similar once those are scaled by the respective peak infection rate. This in turn necessarily assumes that the peak infection for the second wave has already past, which indeed seems to be the case (see Fig. \ref{eight_states_roll}). It also does not consider the effects of vaccinations, new mutant variants of the virus and changes in the travel restriction norms. Nevertheless, use of this `synthetic' training data set enables the ML algorithm to make predictions for the end-time, which is otherwise difficult to do due to the lack of training data sets. 

In conclusion, we note that the epidemic spreading in the SIR model on a two dimensional square lattice can be well predicted by supervised machine learning algorithms. The predictability of the model is sensitive to the site dilution fraction of the model and becomes the highest near the percolation critical point. An optimal condition for predictability can be obtained by tuning the site dilution fraction in the model that minimizes the prediction errors for the first wave of the COVID-19 pandemic. The optimized model can then be used to make predictions of the end-times for the ongoing second wave of the pandemic in different states in India.


\begin{thebibliography}{99}
\bibitem{who}
P.  Zhou,X-L. Yang,X-G. Wang,B. Hu,L. Zhang,W.  Zhang,H-R. Si,Y. Zhu, B. Li, C. Huang, H-D. Chen, J. Chen, Y. Luo, H. Guo, R-D. Jiang, M-Q. Liu, Y. Chen, X-R. Shen, X. Wang, X-S. Zheng, K. Zhao, q-J. Chen, F. Deng, L-L. Liu, B. Yan, F-X. Zhan, Y-Y. Wang, G-F. Xiao, Z-L. Shi, {\it A pneumonia outbreak associatedwith a new coronavirus of probable bat origin}, Nature {\bf 579}, 270 (2020).

\bibitem{chin}
M. Chinazzi, J. Davis, M. Ajelli, C. Gioannini, M. Litvinova, A. Merier,
A. Piontti, K. Mu, L. Rossi, K. Sun, C. Viboud, X. Xiong, H. Yu, 
M. Halloran, I. Longini Jr., A. Vespignani, {\it The effect of travel restrictions on the spread of the 
2019 novel coronavirus (COVID-19) outbreak}, Science {\bf 368}, 395 (2020).


\bibitem{covid5}
J. Dehning, J. Zierenberg, F. P. Spitzner, M. Wibral, J. Pinheiro Neto, M. Wilczek, V. Priesemann, {\it Inferring change points in the spread of COVID-19 reveals the effectiveness of interventions}, Science, DOI: 10.1126/science.abb9789 (2020).

\bibitem{covid6}
M. Kraemer, C.-H. Yang, B. Gutierrez, C.-H. Wu, B. Klein, D. Pigott, Open Covid-19 Data
Working Group, L. du Plessis, N. Faria, R. Li, W. Hanage, J. Brownstein, M. Layan,
A. Vespignani, H. Tian, C. Dye, O. Pybus, S. Scarpino, {\it The effect of human mobility and control measures on the COVID-19 epidemic in China}, Science {\bf 368}, 493 (2020).


\bibitem{covid7}
Z. Ceylan, {\it Estimation of COVID-19 prevalence in Italy, Spain, and France},
Sci. Tot. Env. {\bf 729}, 138817 (2020).

\bibitem{covid8}
C. Anastassopoulou, L. Russo, A. Tsakris, C. Siettos C, {\it Data-based analysis, modelling and forecasting of the COVID-19 outbreak}, PLoS ONE {\bf 15} e0230405 (2020).

\bibitem{covid9}
T. Chakraborty, I. Ghosh, {\it Real-time forecasts and risk assessment of novel coronavirus (COVID-19) cases: A data-driven analysis},
Chaos, Solitons \& Fractals {\bf 135}, 109850 (2020).

\bibitem{ps2}
K. Biswas, A. Khaleque, P. Sen, {\it Covid-19 spread: Reproduction of data and prediction using a SIR model on Euclidean network}, arxiv:2003:07063 (2020).

\bibitem{covid3}
S. Khajanchi, K. Sarkar, J. Mondal, M. Perc, {\it Dynamics of the COVID-19 pandemic in India},
arxiv:2005.06286 (2020).

\bibitem{leo}
{\it Classification and regression trees}, L. Breiman, J. H. Friedman, 
R. A. Olshen, C. J. Stone, CRC Press (1984).

\bibitem{ml_book}
T. Hastie, R. Tibshirani, J. Friedman, {\it The elements of statistical learning: Data mining,
inference and predictions}, Springer – New York, 2001.

\bibitem{labquake}
B. Rouet-Leduc, C. Hullbert, N. Lubbers, K. Barros, C. J. Humphreys, P. A. Johnson,
{\it Machine learning predicts laboratory earthquakes}, Geophys. Res. Lett. {\bf 44}, 9276 (2017).

\bibitem{salm18}
H. Salmenjoki, M. J. Alava, L. Laurson, {\it Machine learning plastic deformation of crystals},
Nat. Commun. {\bf 9}, 5307 (2018).

\bibitem{geophys}
M. van der Baan, C. Jutten, {\it Neural networks in geophysical applications},
Geophysics {\bf 65}, 1032 (2000).



\bibitem{sb}
S. Biswas, D. F. Castellanos, M. Zaiser, {\it Prediction of creep failure time using machine learning},
Sci. Rep. {\bf 10}, 16910 (2020).

\bibitem{sir_o}
W. Kermack, A. McKendrick, {\it A contribution to the mathematical theory of epidemics}, Proc. R. Soc. A {\bf 115}, 700 (1927).


\bibitem{ep1}
N. T. J. Bailey, {\it The mathematical theory of infectious diseases}, 2nd edition, Griffin, London (1975).

\bibitem{ep2}
D. Daley, J. Gani, {\it Epidemic modeling: An introduction}, Cambridge University Press, NY (2005). 


\bibitem{ai_covid1}
V. Chimmula, L. Zhang, {\it Time series forecasting of COVID-19 transmission in Canada using LSTM networks},
Chaos, Solitons \& Fractals {\bf 135}, 109864 (2020). 

\bibitem{ai_covid2}
P. Bedi, S. Dhiman, P. Gole, N. Gupta, V. Jindal, {\it Prediction of COVID-19 Trend in India and Its Four Worst-Affected States Using Modified SEIRD and LSTM Models},
SN Computer Science {\bf 2}, 224 (2021).

\bibitem{syn}
N. Patki, R. Wedge, K. Veeramachaneni, {\it The Synthetic Data Vault}, 2016 IEEE International Conference on Data Science and Advanced Analytics (DSAA), pp. 399-410, (2016).

\bibitem{data}
All data used in this work comes from: https://api.covid19india.org/

\bibitem{testing}
M. J. Binnicker, {\it Challenges and Controversies to Testing for COVID-19}, J. Clinical Microbiology {\bf 58}, e01695-20 (2020).

\bibitem{stau}
D. Stauffer and A. Aharony, {\it Introduction to Percolation Theory},
2nd ed. (Taylor \& Francis, London, 1994).

\bibitem{galam}
K. Malarz, S. Galam, {\it Square-lattice site percolation at increasing ranges of neighbor bonds},
Phys. Rev. E {\bf 71}, 016125 (2005).

\bibitem{city}
M. Batty, P. A. Longley, {\it Fractal cities: a geometry of form and function}, London, Academic Press (1994).


\end{thebibliography}
\end{document}